\documentclass[doublecol]{epl2}

\usepackage{amsmath,amssymb}
\usepackage{xfrac}

\usepackage{graphicx}

\let\originalleft\left
\let\originalright\right
\renewcommand{\left}{\mathopen{}\mathclose\bgroup\originalleft}
\renewcommand{\right}{\aftergroup\egroup\originalright}
\newcommand{\bra}[1]{\ensuremath{\left\langle #1\right|}}
\newcommand{\ket}[1]{\ensuremath{\left|#1\right\rangle}}

\newcommand{\ie}{\emph{i.e.}}

\newcommand{\cf}{cf.}

\title{The optimization topography of exciton transport}

\author{Torsten Scholak\inst{1,2}, Thomas Wellens\inst{1}, and Andreas
Buchleitner\inst{1}}

\institute{
  \inst{1} Physikalisches Institut, Albert-Ludwigs-Universit\"at Freiburg,
  Hermann-Herder-Stra\ss{}e~3, D-79104~Freiburg, Germany, EU\\
  \inst{2} Chemical Physics Theory Group, Department of Chemistry and Center for
  Quantum Information and Quantum Control, University of Toronto, Toronto
  M5S~3H6, Canada
}


\abstract{
  Stunningly large exciton transfer rates in the light harvesting complex of
  photosynthesis, together with recent experimental 2D spectroscopic data, have
  spurred a vivid debate on the possible {\em quantum} origin of such
  efficiency. Here we show that configurations of a random molecular network
  that optimize constructive quantum interference from input to output site
  yield systematically shorter transfer times than classical transport induced
  by ambient dephasing noise.
}

\begin{document}

\maketitle

\section{Introduction}

A growing amount of experimental data
\cite{Engel:2007cr,Panitchayangkoon:2010vn,Fleming:2011tg} suggest that quantum
coherence may be at the origin of the stunning efficiency of exciton transport
in photosynthetic light harvesting, even at ambient temperatures and in a
doubtlessly very noisy environment. The simplest natural structure that exhibits
these surprising properties is the Fenna-Matthews-Olson (FMO) light harvesting
complex of green sulfur bacteria, which consists of seven (or eight
\cite{Olbrich:2010ly}) bacteriochlorophyll molecules arranged in a disordered
network \cite{Fenna:1975ys,Amerongen:2000fk}. Exciton transfer is here mediated
by dipole-dipole coupling between these different molecular sites, and is
associated with a de-excitation of the donor site from some excited to the
ground state, and an excitation of the acceptor site from the ground to some
excited state. The molecular network that mediates the exciton transfer, from
the antenna complex to the reaction center, where the excitation fuels the
organism's chemistry, is embedded into a complicated protein structure, which
seems to provide some structural stiffness, and also defines a nontrivial
spectral structure of the environment, which preserves the coherence on the FMO
network itself, on considerably longer time scales than to be expected for a
white noise environment \cite{Cheng:2009ek}. Since the prevalence of the
coherence effects in widely open systems in contact with high temperature
environments challenges our traditional understanding of what seemed the
restricted realm of quantum mechanics, these highly specialized biological
functional units let us think anew how to control efficient transport in
disordered and noisy systems, possibly by exploiting fundamental quantum
features.

The available experimental evidence, still not always fully consistent
\cite{Engel:2007cr,Panitchayangkoon:2010vn} and vividly debated, is mirrored by
a large variety of theoretical approaches, which distinguish themselves in terms
of the applied methodology as well as of the level of faithfulness of the
modeling of the actual biomolecular object under scrutiny---from advanced
quantum molecular dynamics \cite{Olbrich:2010ly}, over quantum simulations
\cite{Rebentrost:2009vn,Chin:2010fk,Wu:2010vn} based on some effective, seven
site Hamiltonian \cite{Adolphs:2006ve}, Lindblad equations
\cite{Abramavicius:2010fk} or various non-Markovian approaches derived from open
system theory
\cite{Hughes:2009hc,Thorwart:2009fk,Liang:2010nx,Sarovar:2010fk,muelken10}, to
diverse quantum optical models \cite{cai10,alicki11,wuester10} and abstract
statistical treatments \cite{scholak10,Scholak:2011uq}. In essence, there are
four principal lines of argument to explain efficient exciton transport, by (i)
noise-assisted, (ii) non-Markovian or (iii) driving, and (iv) multi-path
interference induced exciton transfer. Given the available experimental data and
the hitherto limited characterization and control of the precise microscopic
Hamiltonian and environment coupling agents that generate the experimentally
observed phenomena, it remains an open question which of these suggested
mechanisms were actually used by evolution to optimize the FMO's functionality.
Since any improvement that provides an evolutionary advantage will be
implemented, it is even not unlikely that all of them are used at some level.
However, in the light of the debate about a possible {\em quantum} enhancement
of transport in the photosynthetic complex, it is highly relevant to understand
the specific role of quantum coherence for these different mechanisms, and to
compare the achievable transfer efficiencies they allow for. It is the present
contribution's purpose to provide such comparison for noise-enhanced and
multi-path-interference transport scenarios. Large scale statistical sampling
will allow us to show that multi-path quantum interference always leads to
better results than noise-assisted, essentially classical transport processes,
though requires additional optimization of the molecular structure. Indeed, this
even holds in the presence of not too strong ambient noise, and thus identifies
yet another scenario where genuine quantum effects may define an evolutionary
advantage.

\section{Random molecular networks}

The noise-enhanced as well as the multi-path-interference scenario start out
from the same structural elements that are unambiguously given by experimental
observations: Exciton transfer occurs across a random molecular network with
local sites effectively modeled as electronic two-level systems, and further
background degrees of freedom, which may possibly exert some effective driving
\cite{cai10}, can in principle be accounted for by some non-Markovian
environment coupling \cite{Hughes:2009hc}. Experimentally observed exciton
cross-terms \cite{Engel:2007cr}, though with quite some experimental scatter
\cite{Panitchayangkoon:2010vn}, suggest excitonic coherence times which exceed
the exciton transfer time between antenna and reaction center by a factor two to
five \cite{Engel:2007cr}, even at ambient temperatures
\cite{Panitchayangkoon:2010vn}. The presently best available effective
Hamiltonian, which is inferred from experimental data and advanced quantum
chemical model calculations predicts a strong suppression of strictly coherent
transport across the FMO complex (approx. $5\ \%$ transfer efficiency
\cite{scholak10,Scholak:2011uq}), from the input to the output site, due to
predominantly destructive multi-path interference upon transmission.

However, exciton transfer in the FMO complex occurs with efficiencies larger
than $95\ \%$ \cite{Cheng:2009ek}, and there are essentially two alternative
scenarios which can explain this discrepancy between the experimental data and
the best microscopic Hamiltonian presently available. Since quantum coherence
can be destroyed by ambient noise, and since suppressed transport under purely
coherent dynamics can only be due to destructive interference effects, it is
very natural to argue in favor of noise-induced transport \cite{arndt91}. Since,
however, destructive multi-path interference is known to be very sensitive with
respect to changes of the boundary conditions and/or the Hamiltonian's coupling
matrix elements \cite{Kramer:1993yb}, and since even the best available model
Hamiltonian for the FMO complex is garnished with appreciable uncertainties for
its individual entries \cite{Adolphs:2006ve}, one may equally well argue in
favor of {\em constructive} multi-path interference as the observed efficiency's
cause. This even more so since all experimental data that so far lend support
for the coherent transport hypothesis are obtained from bulk measurements rather
than from single molecule spectroscopy, and therefore might mask much longer
coherence times by inhomogeneous broadening effects \cite{Fleming:2011tg}.

Since both alternative explanations allow to predict large transfer
efficiencies, let us have a closer look at the respective key mechanisms. We
model the energy conserving, unitary dynamics of a single
excitation injected into a molecular network alike the FMO complex by the
Hamiltonian
\begin{align}
  H &= \sum_{i \neq j = 1}^7 v_{i,j} \ket{i}\bra{j}\, ,
  \label{ham}
\end{align}
with $\ket{i}$ and $\ket{j}$ the electronic states where the excitation is
localized at the individual molecular sites $i$ and $j$, respectively. We assume
that initially only one site, ``in'', is excited, which is identified with the
first site of the network. The sites No.~$2$ to $6$ are referred to as the
``intermediates''. The seventh and last site is the designated output site,
``out'', where we add an energy sink, such as to mimic the irreversible exciton
absorption at the reaction center. We couple each of the seven molecular sites
to a private (\ie{}, there is no inter-site communication through the
environment) dephasing environment. Sink and dephasing induce some
irreversibility on the FMO degrees of freedom, which we incorporate by the
Lindblad terms
\begin{align}
  L_{\mathrm{sink}}(\varrho) &= \Gamma
  \bigl(\ket{0}\bra{\mathrm{out}}\varrho\ket{\mathrm{out}}\bra{0} - \tfrac{1}{2}
  \{\ket{\mathrm{out}}\bra{\mathrm{out}}, \varrho\}\bigr),
  \label{lind_sink}
\end{align}
where $\ket{0}$ and $\{,\}$ are the ground state of the molecular
network and an anticommutator, respectively, and
\begin{align}
  L_{\mathrm{deph}}(\varrho) &= - 4 \gamma \sum_{i \neq j = 1}^7
  \ket{i}\bra{i}\varrho\ket{j}\bra{j}
  \label{lind_deph}
\end{align}
into the effective evolution equation of the excitonic state on the
network,
\begin{align}
  \dot{\varrho}(t) &= - \mathrm{i} \, [H, \varrho(t)] +
  L_{\mathrm{sink}}(\varrho(t)) + L_{\mathrm{deph}}(\varrho(t))
  \label{lind_FMO}
\end{align}
(where we have set $\hbar = 1$ for convenience). To obtain a robust
comparison of the different transport mechanisms, we statistically sample the
transport efficiency over different realizations of $H$, by random sampling over
the positions of all intermediate sites, within a sphere with input and output
site placed on its north and south pole, respectively. Random positions
$\vec{r}_j$ translate into random realizations of $H$ by defining the coupling
matrix elements as a function of the intersite distances,
\begin{align}
  v_{i,j} &= \alpha \, r_{i,j}^{-3}
  \label{random_v}
\end{align}
with some constant $\alpha$, and $r_{i,j}=|\vec{r}_i-\vec{r}_j|$. By choosing
random positions for the intermediate sites, we in some way take into account
the experimental uncertainty concerning the actual FMO Hamiltonian
\cite{Adolphs:2006ve}. Let us stress, however, that we do not present our model
as a realistic description of a particular experiment. On the contrary, we are
rather interested in general properties of transport, as they will appear if we
allow for a large variety of possible configurations, without imposing, from the
very beginning, strict experimental boundary conditions.

The model discussed above has three intrinsic time scales which will largely
determine the expected transfer efficiencies---the direct exciton transfer time
\begin{align}
  T &= \frac{\pi}{2 |v_{\mathrm{in},\mathrm{out}}|}
  \label{eq:T}
\end{align}
between input and output site, in the absence of all intermediate sites, the
local dephasing rate $4\gamma$ (identical for all sites), and the sink
dissipation rate $\Gamma$, which we will fix at the value $\Gamma = 10/T$ in the
following \footnote{This implies one incoherent event on the time scale $T/10$,
hence a rather efficient drain. We have used this time scale as an efficiency
benchmark in earlier work \cite{scholak10,Scholak:2011uq}.}. Randomly placed
intermediate sites between input and output help to enhance or suppress coherent
transport \cite{scholak10,Scholak:2011uq}, by larger coupling strengths between
closer sites and appropriate phase relationships. The dephasing rate defines the
time scale $\mathfrak{T}_{\mathrm{deph}}=(4\gamma)^{-1}$ on which such phase
relationships can have a bearing on the overall transport behavior, and the sink
dissipation rate defines an optimal time scale on which population has to be
delivered to the output site, to make it immediately available for the reaction
center.

\section{Statistics of transport efficiency}

\begin{figure*}
  \hfill
  (a)\raisebox{-26.04681pt}{\includegraphics{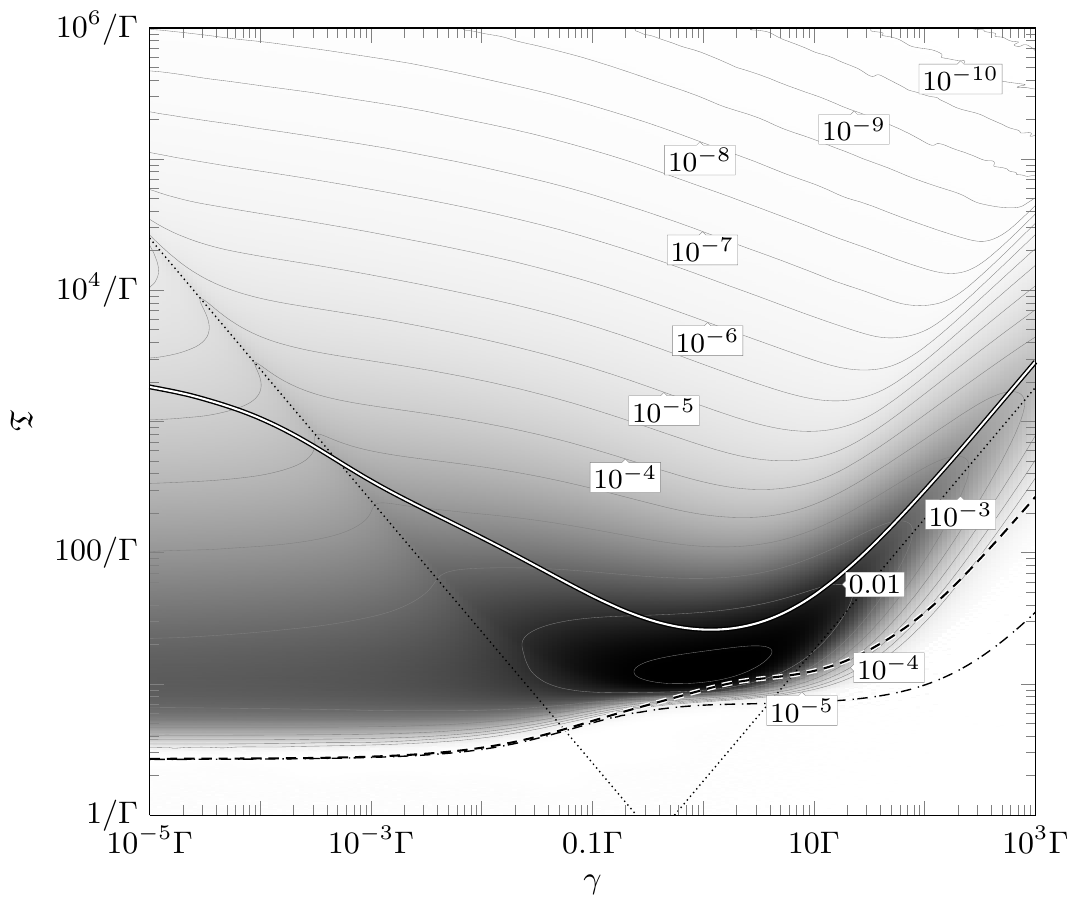}}
  \hfill
  (b)\raisebox{-28.42184pt}{\includegraphics{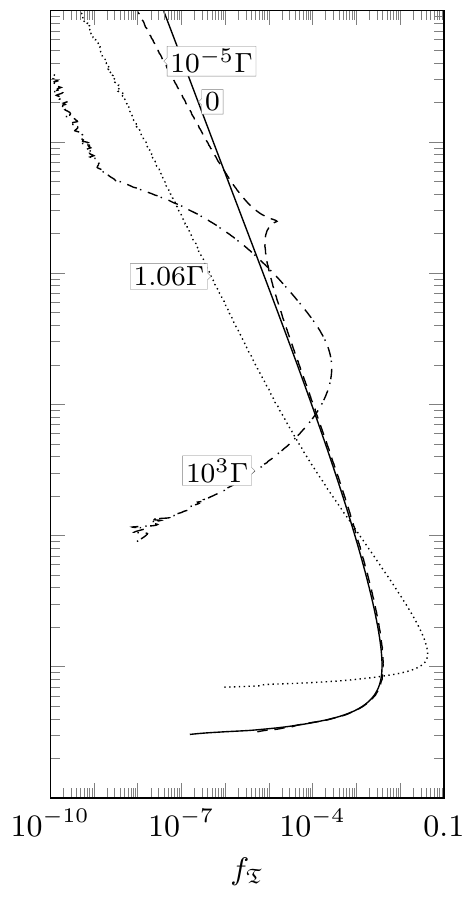}}
  \hfill
  \caption{(a) Probability density $f_{\mathfrak{T}}$ of the average excitation
  transfer time $\mathfrak{T}$, eq.~\eqref{eq:mathfrakT}, for $N=7$ molecular
  sites and sink rate $\Gamma=10/T$, as a function of the dephasing rate
  $\gamma$. The two dotted, diagonal lines are given by the dephasing time
  $\mathfrak{T}_{\mathrm{deph}}= (4 \gamma)^{-1}$, and by an approximate Zeno
  time $\mathfrak{T}_{\mathrm{Zeno}}\propto\gamma$, respectively. On time scales
  $\mathfrak{T}>\mathfrak{T}_{\mathrm{deph}}$, the purity of the excitonic state
  on the molecular network has dropped to its minimum value, hence the transport
  is essentially classical. The white line shows the median
  $\tilde{\mathfrak{T}}$, the dot-dashed line the minimum transfer time, and the
  dashed line the transfer time of a configuration that has been optimized for
  $\gamma = 0$. (b) Height profiles of the probability densities for fixed
  dephasing rates $\gamma =0$ (solid line), $10^{-5}\Gamma $ (dashed line),
  $1.06\Gamma $ (dotted line), $10^3\Gamma $ (dash-dotted line).}
  \label{topography}
\end{figure*}

To compare the transport efficiency provided by different molecular
conformations and different dephasing rates, we define the average transfer time
\begin{align}
  \mathfrak{T} &= \Gamma\int_0^\infty t \, p_{\mathrm{out}}(t) \, \mathrm{d}t \,
  ,
  \label{eq:mathfrakT}
\end{align}
which is the expectation value of the time required to transfer the excitation
to the sink, determined by the population $p_{\mathrm{out}}(t) = \bra{0}
\dot{\varrho}(t) \ket{0} / \Gamma$ of the output site. We see here that the
ground state $\ket{0}$ can only be populated by delivering the exciton from the
output site to the sink, with rate $\Gamma$, \cf{} also eq.~\eqref{lind_sink}.
The shorter the transfer time, the more efficient the transport \footnote{Note
that $\mathfrak{T}$ is a reasonable efficiency quantifier if efficiency is
qualified as rapid and irreversible excitation transfer to the sink, but that it
does not distinguish quantum from classical transport efficiencies, since it
integrates over all times. It is however evident that the definition of any
efficiency quantifier is a matter of pragmatic choice rather than of principle,
and that all such quantifiers call for a careful interpretation}.

We are now prepared for a statistical analysis of exciton transfer times across
a molecular network alike the FMO complex, to assess the potential role of
coherent vs. noise-assisted transport mechanisms to steer transport
efficiencies. In order to draw a landscape of the exciton transport efficiency
in molecular networks as modeled by eq.~\eqref{lind_FMO}, we sample over fifty
million random and distinct conformations. For each conformation in the ensemble
the computational procedure involves the following tasks: First, the positions
of the five intermediate molecular sites are randomized, which are then used to
populate the matrix entries of the Hamiltonian $H$ according to
eq.~\eqref{random_v}. For values of $\gamma$ from $10^{-5} \Gamma$ to $10^3
\Gamma$ in 200 logarithmic steps, \ie{} for essentially vanishing to very strong
dephasing rates, the master equation, eq.~\eqref{lind_FMO}, is solved via exact
numerical diagonalization and the transfer times $\mathfrak{T}$ are calculated.
The last step employs logarithmic data binning to record the transfer time
histogram $f_{\mathfrak{T}}$, shown in Fig.~\ref{topography}, as a function of
$\gamma$. In the left density plot of the figure, the grey scale represents the
probability density of configurations giving rise to a certain transfer time,
under a given dephasing rate. Configurations above the left hand, decreasing
dotted line correspond to transfer times {\em longer} than
$\mathfrak{T}_{\mathrm{deph}}$, \ie{} the exciton transfer is here due to {\em
classical transport} across the network. For even larger dephasing rates,
transport is ``frozen'' due to a Zeno-like projection mechanism
($\mathfrak{T}_{\mathrm{Zeno}} \propto \gamma$, as indicated by the increasing
dotted line), while configurations below the left hand dotted line achieve
efficient transfer on time scales {\em shorter} than
$\mathfrak{T}_{\mathrm{deph}}$. The latter thus mediate {\em coherent exciton
transfer} and are the only ones which are eligible for claiming an unambiguous
{\em quantum enhancement} of exciton transfer in the FMO complex. Indeed, the
shortest transfer times are observed for a finite subset of these
configurations, on the lower left hand side of Fig.~\ref{topography}. If we
furthermore optimize the molecular configuration in the absence of any
dephasing, for the same sink dissipation, and then expose this optimal
configuration to finite dephasing rates, we obtain the dashed curve in
Fig.~\ref{topography}, which exhibits exciton transfer times {\em shorter} than
the classical value, associated with the highest density of configurations in
the plot, by at least a factor two, for almost all dephasing rates that allow
for coherent transfer. Note that this optimal configuration still yields very
efficient excitation transfer even for higher dephasing rates, which suggests
that the optimized coherent transport on short time scales imparts an initial
advantage, even when classical activation takes over on long time scales.

\section{Discussion}

The optimization landscape thus provides a clear picture of the possible
strategies to achieve efficient transfer across an FMO like, disordered network:
The large majority of randomly sampled configurations requires the {\em
assistance of noise} to allow for efficient transport, which, however, will be
{\em classical} in nature, and will occur on time scales clearly longer than
$\mathfrak{T}_{\mathrm{deph}}$. Indeed, the clustering of all the configurations
associated with this transport mechanism around $\mathfrak{T} \simeq 10/\Gamma
=T$ and $\gamma \simeq \Gamma$ in Fig.~\ref{topography} expresses an approximate
matching condition of dephasing and sink dissipation rates, which is an
intrinsically classical, statistical synchronization phenomenon. Excitation
transfer here stems from the noise induced destruction of quantum coherence, is
induced on time scales {\em longer} than the system inherent coherence times,
and is essentially {\em independent} of the microscopic Hamiltonian which
generates the coherent dynamics on the network (hence the large density of
configurations at these transfer times and dephasing rates). Under an
evolutionary perspective, this non-selectivity with respect to the microscopic
coupling structure can therefore neither define an evolutionary advantage by
optimizing the molecular structure---most structures will do, as also
highlighted by the pronounced minimum of the median of the distribution (white
curve) at $\gamma\simeq\Gamma$.

In contrast, Fig.~\ref{topography} also shows that a small but finite
sub-ensemble of {\em optimal configurations} mediates {\em efficient quantum
transport} of the excitation from input to output, faster than the
noise-assisted, classical transport, and due to constructive quantum
interference upon transmission, which prevails in the presence of noise,
actually in the entire range of dephasing rates considered in
Fig.~\ref{topography}. If quantum mechanics is at the origin of the
experimentally observed exciton transfer efficiencies in the FMO complex, it
therefore must stem from such type of optimal molecular configurations, or
otherwise ought to be induced by non-Markovian \cite{Thorwart:2009fk} or driving
effects \cite{cai10}. Since the optimal transfer rates mediated by such optimal
configurations define rare events in the statistical sample represented in
Fig.~\ref{topography}, it is conceivable that they define an evolutionary
advantage which was hardwired by nature. The cusp-like structure that emerges
right along the line defined by the dephasing time
$\mathfrak{T}_{\mathrm{deph}}$ highlights molecular configurations which exhibit
an eigenstate of $H$ localized on the site ``in'' and on another site $j \neq
\text{``out''}$. For these configurations, coherent transmission is suppressed
by destructive interference, and hence noise is required to assist the exciton's
delivery at the sink, with transfer times $\mathfrak{T} =
\mathfrak{T}_{\mathrm{deph}}$.

Let us conclude with a short discussion of the actual structure of optimal
configurations in the coherent regime: When investigating the molecular
conformations which exhibit transfer times shorter than the dephasing time
$\mathfrak{T}_{\mathrm{deph}}$ in Fig.~\ref{topography}, these show a wide
variability as regards their dimensionality and symmetry properties. Minimal
transfer times are achieved by near to one dimensional structures, while fully
three dimensional and prima facie disordered structures, that mediate optimal
coherent transport on closed molecular networks with vanishing sink dissipation,
achieve slightly longer transfer times $\mathfrak{T}$, though still
significantly shorter than the dephasing time. Since Fig.~\ref{topography} is
obtained for a relatively large sink dissipation rate, which corresponds to a
significant change of boundary conditions as compared to the closed network,
this is not surprising, though raises the question for the actual optimality
conditions favored by nature: Besides rapid transfer, also robustness and
multifunctionality issues define further constraints which go far beyond the
scope of our present model considerations, but define a beautiful and
fascinating area for further research. Therefore, it remains an open
question, whether nature indeed employs quantum coherence to benefit energy
transport in photosynthesis by having evolved molecular conformations
corresponding to the optimal configurations discovered here. It appears that due
to the strong sensitivity of the transport efficiency on the particular
realization of disorder this question can only be answered by a highly precise
measurement of the electronic Hamiltonian with an accuracy beyond what is
accessible today. Finally, regardless of whether the near-to-perfect efficiency
of the FMO complex is truly caused by quantum coherence or not, the existence
of optimal configurations achieving maximum performance due to constructive
quantum interference will certainly spur the design of new experiments and, in
the long run, advanced devices such as a new generation of organic solar cells
that utilize the beneficial aspects of quantum coherence.


\end{document}